\documentclass{pasj00}
\begin{document}
\SetRunningHead{N. Narita et al.}{Transmission Spectroscopy of the HD~209458b with Subaru HDS}
\Received{2005/03/07}
\Accepted{2005/04/25}

\title{Subaru HDS Transmission Spectroscopy\\of the Transiting Extrasolar Planet HD~209458b}


\author{
Norio \textsc{Narita},\altaffilmark{1}
Yasushi \textsc{Suto},\altaffilmark{1}
Joshua N.\ \textsc{Winn},\altaffilmark{2}$^{*}$
Edwin L.\ \textsc{Turner},\altaffilmark{1,3}
\\
Wako \textsc{Aoki},\altaffilmark{4}
Christopher J. \textsc{Leigh},\altaffilmark{5}
Bun'ei \textsc{Sato},\altaffilmark{6}
Motohide \textsc{Tamura},\altaffilmark{4} and
Toru \textsc{Yamada},\altaffilmark{4}
}

\bigskip

\altaffiltext{1}{Department of Physics, School of Science, The
University of Tokyo, Tokyo 113-0033}
\email{narita@utap.phys.s.u-tokyo.ac.jp, suto@phys.s.u-tokyo.ac.jp}

\altaffiltext{2}{Harvard-Smithsonian Center for Astrophysics,\\
60 Garden St.\ MS--51, Cambridge, MA 02138, USA}
\email{jwinn@cfa.harvard.edu}

\altaffiltext{3}{Princeton University Observatory, Peyton Hall,
Princeton, NJ 08544, USA} \email{elt@astro.princeton.edu}

\altaffiltext{4}{National Astronomical Observatory of Japan, 2--21--1,
Mitaka, Tokyo 181--8588} \email{aoki.wako@nao.ac.jp, 
tamuramt@cc.nao.ac.jp\\yamada@optik.mtk.nao.ac.jp}

\altaffiltext{5}{Liverpool John Moore's University, Astrophysics
Research Institute,\\ Twelve Quays House, Egerton Wharf,
Birkenhead, CH41, 1LD, U.K.} \email{cjl@astro.livjm.ac.uk}

\altaffiltext{6}{Graduate School of Science and Technology, Kobe
University, 1-1 Rokkodai, Nada, Kobe 657-8501}
\email{satobn@kobe-u.ac.jp}


\maketitle

\begin{abstract}

We have searched for absorption in several common atomic species due
to the atmosphere or exosphere of the transiting extrasolar planet
HD~209458b, using high precision optical spectra obtained with the
Subaru High Dispersion Spectrograph (HDS).  Previously we
reported an upper limit on H$\alpha$ absorption of 0.1\% (3$\sigma$)
within a 5.1~\AA\ band.  Using the same procedure, we now report upper
limits on absorption due to the optical transitions of Na D, Li,
H$\alpha$, H$\beta$, H$\gamma$, Fe, and Ca. The 3$\sigma$ upper limit
for each transition is approximately 1\% within a 0.3~\AA\ band (the
core of the line), and a few tenths of a per cent within a 2~\AA\ band
(the full line width). The wide-band results are close to the expected
limit due to photon-counting (Poisson) statistics, although in the
narrow-band case we have encountered unexplained systematic errors at
a few times the Poisson level. These results are consistent with all
previously reported detections and upper limits, but are significantly
more sensitive.

\end{abstract}
\footnotetext[*]{Hubble Fellow.}

\section{Introduction}

The first star other than the Sun for which a planetary transit was
detected was HD~209458.  The initial discovery by
\citet{2000ApJ...529L..45C} and \citet{2000ApJ...529L..41H}, along
with subsequent observations of higher accuracy by
\citet{2001ApJ...552..699B} and \citet{2000ApJ...540L..45J}, allowed
measurements of the orbital inclination ($i = 86.\!\!^{\circ}1$) and
the planetary radius ($R = 1.34$~$R_{\rm Jup}$).  By combining these
results with the stellar radial velocity data, it was shown that the
companion has a mass comparable to Jupiter ($M = 0.69$~$M_{\rm Jup}$)
and the density of a gas giant planet ($\rho \sim 0.4$ g cm$^{-3}$).

Observations of planetary transits also provide an opportunity to
study the atmospheric properties of close-in giant extrasolar planets
(also known as ``hot Jupiters'').  During occultations, a small
fraction of the stellar disk is blocked from view, and an even smaller
fraction of the starlight is transmitted through the thin,
partially-transparent portion of the planetary atmosphere and
exosphere. Thus, by obtaining spectra with high spectral resolution and high
signal-to-noise ratio, and comparing the spectra taken in and out of
transit, one can detect absorption or emission features due to the
planetary atmosphere or exosphere. This methodology is referred to as
``transmission spectroscopy.''

Several studies of HD~209458 using transmission spectroscopy have
previously been reported. Most have provided upper limits on
particular absorption features. The first positive detection of an
extrasolar planetary atmospheric absorption feature was made by
\citet{2002ApJ...568..377C}. Their motivation was to confirm the
prediction by \citet{2000ApJ...537..916S} (and later,
\cite{2001ApJ...553.1006B}; \cite{2001ApJ...560..413H}) of a
relatively strong absorption feature in the sodium doublet lines.
They observed HD~209458 with the Space Telescope Imaging Spectrograph
(STIS) onboard the Hubble Space Telescope (HST), and found an
additional $0.023\pm 0.006$\% absorption during transits within a band
pass centered on the sodium doublet lines ($5893 \pm 6$~\AA). However,
this additional absorption was significantly weaker than that
predicted. Later, detection of excess absorption in the Ly$\alpha$
(1216~\AA) transition of neutral hydrogen was reported by
\citet{2003Natur.422..143V}, also based on HST/STIS spectra. These
authors found an amazingly strong additional Ly$\alpha$ absorption of
$15\pm 4$\% during transits, implying that the absorbing H~{\sc i} gas
extends over several planetary radii, and thus should be interpreted as
an exosphere rather than an atmosphere. \citet{2004ApJ...604L..69V}
subsequently reported 2--3 $\sigma$ detections of oxygen (O~{\sc i}) and
carbon (C~{\sc ii}) using HST/STIS. Interestingly, among the seven known
cases of transiting hot Jupiters
\citep{2002AcA....52....1U, 2002AcA....52..115U, 2003AcA....53..133U,
2004ApJ...613L.153A}, HD~209458b has both the smallest
mass and the largest radius, and thus an anomalously low density. This
fact makes it yet more interesting to study the very extended
exosphere of HD~209458b.

These reports of an extended exosphere around HD~209458b have inspired
searches for additional absorption features.  However, in spite of the
successful detections via space-based observations, no data obtained
from the ground has yet confirmed the detections or discovered any new
components. Upper limits on planetary exospheric absorption features
of various components in the optical region were reported by
\citet{2000PASP..112.1421B} based on Keck I/HIRES data. They obtained
transmission spectra of HD~209458 (H$\beta$, H$\gamma$, Fe) and 51
Pegasi (Li, H$\alpha$, Na D1, and D2), and placed upper limits of
3--20\% on additional absorption during transits in a 0.3~\AA\
bandwidth centered on the cores of the relevant stellar lines.
\citet{2001A&A...371..260M} also searched for transit effects 
(especially for ionized species) in the
optical band with the VLT/UVES.  They did not set upper limits on
any specific atomic lines, but estimated a detection limit of order 1\%
within any 0.2~\AA\ band covered by the observations.
Near-infrared features have also been examined to set more stringent
upper limits on He, CO, H$_2$O, and CH$_4$ 
(e.g. \cite{2002PASP..114..826B}; \cite{2002AAS...201.4604H};
\cite{2003A&A...405..341M}; \cite{2005ApJ...622.1149D}).

In this paper we report on the results of searches for excess Na (D1, D2),
Li, H$\beta$, H$\gamma$, Fe and Ca absorption in optical spectra of
HD~209458 during transits, based on Subaru High Dispersion
Spectrograph (HDS) data acquired in 2002 October. Compared with the
previous investigations, the HDS data cover a larger range of
wavelengths (covering the entire optical band) and have a higher
signal-to-noise ratio.
Moreover, this data set has a substantial
advantage of covering a wide range of orbital phases, namely before,
during and after the transit in a single night.  The major obstacles
to ground-based transmission spectroscopy are time-dependent effects
due either to instrumental changes caused by variations in position,
temperature et cetera, or to telluric effects such as seeing,
airmass, and atmospheric absorption.  All such effects are most
effectively monitored, interpolated and removed if the data are
obtained on a single night and if an entire transit is observed.
We employed the same reduction and analysis techniques
used and described in detail in \citet{2004PASJ...56..655W} (hereafter
Paper I) in which we set an upper limit of 0.1\% (3$\sigma$)
additional absorption in a 5.1~\AA\ band centered on H$\alpha$.

This paper is organized as follows.  Section 2 describes the
observations, summarizes the reduction method and points out some
small differences between the techniques employed in the present paper
and Paper I.  We present results for the selected absorption lines in
section 3 and discuss systematic effects in section 4.  Section 5
provides a brief summary and interpretation of the results.

\section{Observations and Reduction Method}

\begin{figure}[thb]
 \begin{center}
  \FigureFile(65mm,65mm){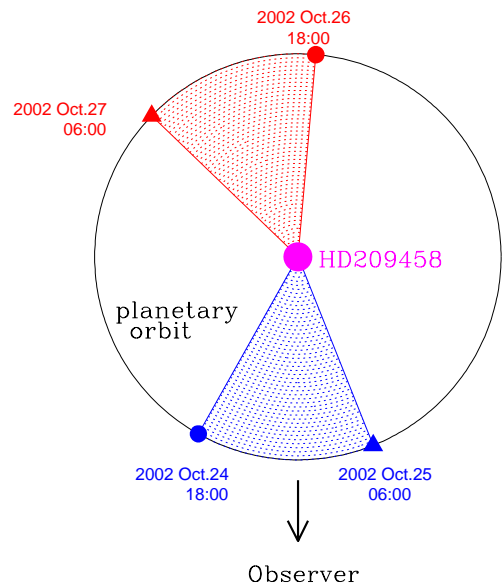} \caption{Phase of the Subaru observations (times given in HST).}
  \label{fig1}
 \end{center}
\end{figure}

We observed HD~209458 on the nights of HST~2002~October~24 and 26 (25
and 27 in UT) with the Subaru HDS \citep{2002PASJ...54..855N}.
The planet was predicted to transit its parent star near the middle of
the night of HST October~24 (see figure~\ref{fig1}), according to the
ephemeris of \citet{2001ApJ...552..699B}.  In order to determine
accurate radial velocities of the star and to verify the ephemeris, we
obtained a spectrum with an Iodine Cell at both the start and the end
of this series of exposures, and confirmed our expectation that the
transit took place when predicted.  We obtained 30 spectra of
HD~209458, and 7 spectra of the rapidly rotating B5~Vn star HD~42545
\citep{2002ApJ...573..359A} in order to evaluate the telluric
spectrum.
The wavelength coverage was 4100~\AA\ $< \lambda <$ 6800~\AA\ with a
spectral resolution of $R \approx 45000$ (0.9~km~s$^{-1}$ per pixel). 
The typical exposure time was 500 seconds with seeing of
$\sim 0\farcs6$, through air masses ranging from 1.0 to 1.9.
The resulting spectra had a typical signal-to-noise ratio of 
$\approx 350$ per pixel.

First, we processed the frames with standard IRAF\footnote{The Image
Reduction and Analysis Facility (IRAF) is distributed by the U.S.\
National Optical Astronomy Observatories, which are operated by the
Association of Universities for Research in Astronomy, Inc., under
cooperative agreement with the National Science Foundation.}
procedures and extracted one-dimensional spectra. Next, we applied an
analysis method described at length in Paper I to correct for
substantial time-dependent variations of the instrumental blaze
function. However, a slight modification of the Paper I procedure was
employed. In Paper I, we used all 4100 pixels of each spectral order
to calculate the best-fitting parameters of a smooth function that
describes the instrumental variations; however, we noted that this
procedure acts to dilute any real signal.  An artificially injected
signal of 0.1\% absorption was recovered with a strength of only
0.057\%. We have since found that this problem is significantly
reduced if the domain of pixels used in the fitting procedure is
restricted to approximately 1000 pixels in the vicinity of the target
line. This modification increases the sensitivity of the analysis and
also provides a modestly improved correction for instrumental
variations (see section 4).

Next, a template spectrum $T (\lambda)$ is created from all 30
individual exposures using the above correction method 
(see Paper I for further details).
A time series of residual spectra are generated by subtracting the suitably
matched template $T_{\rm m} (\lambda)$ from each spectrum $S (\lambda,t)$;
thus the template spectrum determines the zero-level for the residuals,
and therefore we use all of the 30 spectra (including in-transit spectra)
so as to create an optimum template spectrum.
Then, we examine the time
variation of each selected absorption line in order to search for
possible additional absorption in the transit phase.  Specifically, we
compute a time series of fractional difference in flux between the
spectrum and the matched template within a smoothing width $\Delta
\lambda$ according to
\begin{equation}
\delta_{\rm A}(t) =
   \frac{\displaystyle \int_{\lambda_0 - \Delta \lambda/2}^{\lambda_0 + \Delta \lambda/2} d\lambda 
\hspace{0.1in} \left[ S(\lambda, t) - T_{\rm m}(\lambda) \right] }
        {\displaystyle \int_{\lambda_0 - \Delta \lambda/2}^{\lambda_0 + \Delta \lambda/2} d\lambda 
\hspace{0.1in} T_{\rm m}(\lambda) },
\end{equation}
where ${\rm A}$ indicates each target absorption line. 
We refer to such a time variation of the fractional difference as a
``difference light curve''.

To place a quantitative limit on additional absorption, we
calculate $\bar{\delta}_{\rm in,A}$ and $\sigma_{\rm in,A}$:
\begin{eqnarray}
\bar{\delta}_{\rm in,A} &\equiv& \frac{1}{N_{\rm in}} \sum_{t \rm =in} \delta_{\rm A}(t) \, ,\\
\sigma_{\rm in,A} &\equiv& \sqrt{\frac{1}{N_{\rm in}} \sum_{t \rm =in} \left[ \bar{\delta}_{\rm in,A} - \delta_{\rm A}(t) \right]^2} \, ,
\end{eqnarray}
which are the mean value and the standard deviation of $\delta_{\rm
A}(t)$ for the spectra taken in the transit (between second and third
contacts, namely within $\pm$ 60 minutes from center of the transit).
Similarly, we compute $\bar{\delta}_{\rm out,A}$ and $\sigma_{\rm
out,A}$ from the spectra taken out of the transit (before first
contact or after fourth contact: beyond $\pm$ 92 minutes from center
of the transit).
\footnote{To compute the transit times, we
use an ephemeris defined by J.D.~(mid-transit) = 2451659.93675 
\citep{2001ApJ...552..699B} and the most precise published period, 
$P=3.524739$~d \citep{2000A&A...355..295R}.} 
The numbers of spectra in each phase are
$N_{\rm in} = N_{\rm out} = 12$.

Finally, we compute
\begin{eqnarray}
\bar{\delta}_{\rm A} &\equiv& \bar{\delta}_{\rm in,A} - \bar{\delta}_{\rm out,A} \, ,\\
\sigma_{\rm A} &\equiv& \sqrt{\frac{N_{\rm in} \sigma_{\rm in,A}^2 + N_{\rm out} \sigma_{\rm out,A}^2}{N_{\rm in} + N_{\rm out}}} = \sqrt{\frac{\sigma_{\rm in,A}^2 + \sigma_{\rm out,A}^2}{2}} \, .
\end{eqnarray}
Note that $\bar{\delta}_{\rm A}$ would be negative if there were
additional absorption during transit.

In the following sections, we express the results as $\bar{\delta}_{\rm
A} \pm \sigma_{\rm A}$.  Where no additional absorption feature
related to transit are detected (i.e., $|\bar{\delta}_{\rm A}| < 
\sigma_{\rm A}$), we quote upper limits as 3$\sigma_{\rm A}$.

\section{Results}

Since the HDS spectra cover a wide wavelength region in optical band
($4100 --- 6800$~\AA), there are many absorption lines available for
inspection.  We concentrate on the following transitions for detailed
study: 1) the sodium and lithium lines, because additional absorptions
at these lines are theoretically predicted by
\citet{2000ApJ...537..916S}; 2) neutral hydrogen lines including
H$\alpha$, H$\beta$, and H$\gamma$, since an extended hydrogen exosphere
was reported by \citet{2003Natur.422..143V}; and 3) the strong stellar
lines of Ca and Fe.  Another reason to examine the sodium doublet
lines in detail is to attempt to confirm the report by
\citet{2002ApJ...568..377C} of additional $0.023\pm 0.006$\%
absorption in the $5893 \pm 6$~\AA\ band.

In Paper~I, we paid particular attention to the results for a
bandwidth of 5.1~\AA, in order to facilitate a comparison with the band
used by \citet{2003Natur.422..143V}. In this work we generally
consider two different bandwidths: a relatively narrow band of 0.3~\AA\
[so as to facilitate comparison with the previous results of
\citet{2000PASP..112.1421B} and \citet{2001A&A...371..260M}], and a
relatively wide band of 2~\AA\ to cover the full width of target
absorption lines and provide more sensitivity.

\subsection{Calculation of Difference Spectra}

\begin{figure*}[tbh]
 \begin{center} \FigureFile(160mm,250mm){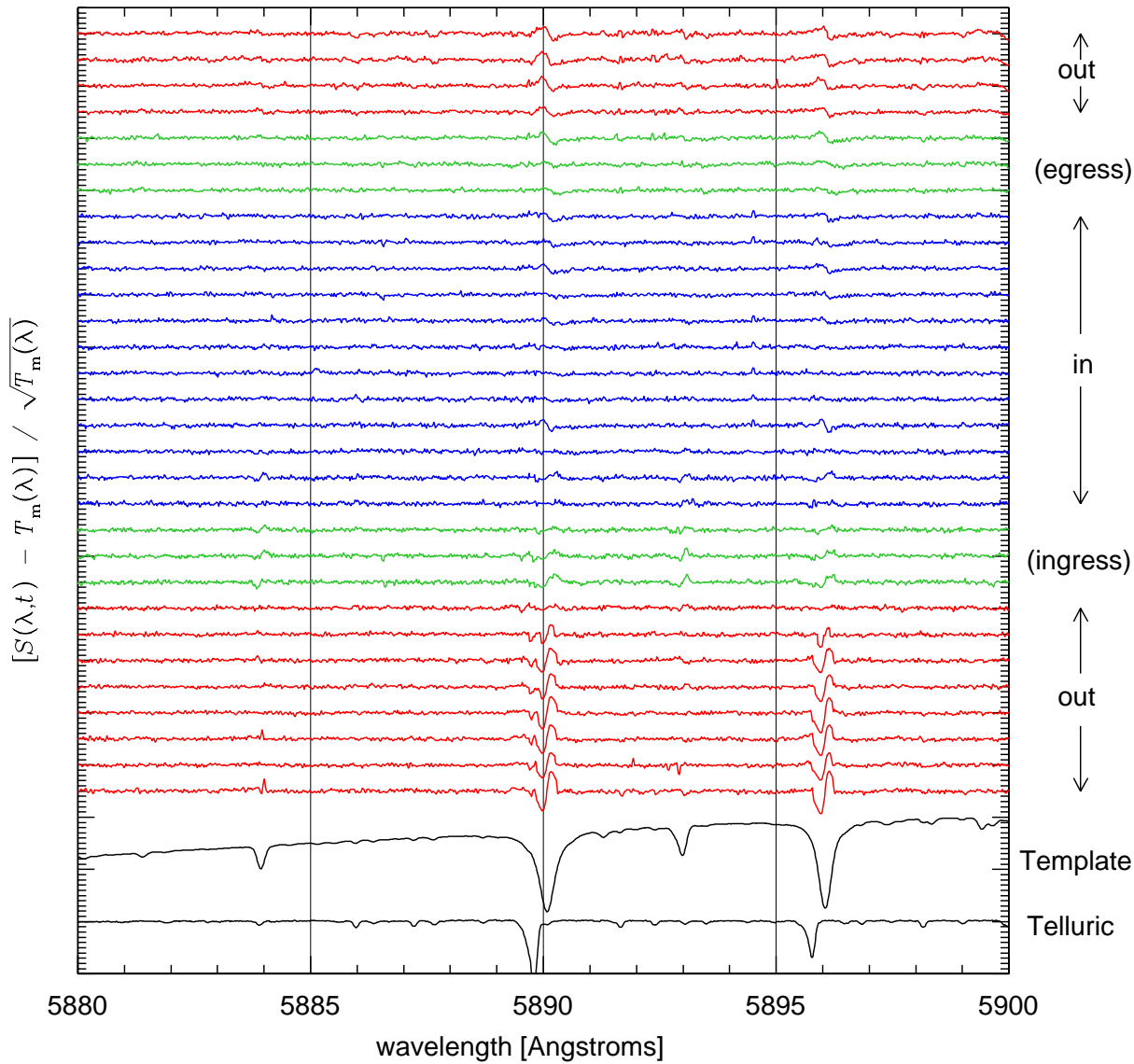} \caption{Difference 
  between each spectrum and its matched template for a
  wavelength range in the vicinity of the sodium doublet lines. The
  spectra are ordered in time from bottom to top. The corresponding
  phases of the planetary transit are noted on the right side of the
  figure. The spacing between minor ticks represents two Poisson
  deviations. The bottom spectrum shows the normalized template
  spectrum of the rapidly rotating star HD~42545, containing telluric
  features and interstellar sodium absorption. The second spectrum
  shows the overall template spectrum of HD~209458. Note that we do
  not correct to the laboratory frame.} \label{fig2}
  \end{center}
\end{figure*}

We express the difference between each spectrum and the matched
template in units of the expected standard deviation due to Poisson
statistics,
\begin{equation}
\frac{S (\lambda) - T_{\rm m} (\lambda)}{\sqrt{T_{\rm m} (\lambda)}},
\end{equation}
and refer to this quantity as a ``difference spectrum''. As an
example, the difference spectra in the vicinity of the sodium
resonance doublet are shown in figure~\ref{fig2}. The template
spectrum and the telluric spectrum are plotted at the bottom of the
series.  To estimate the telluric spectrum, we created a template
spectrum for HD~42545 in the same way as for HD~209458 and normalized
it appropriately for visual reference.

The template spectrum shows some features that are similar to those
seen in the estimated telluric spectrum.  However, the latter spectrum
is in reality not just the telluric spectrum; it is also affected by
interstellar sodium absorption. Moreover, the sodium D2 line at
5890~\AA~is blended with strong telluric vapor absorption (e.g.,
\cite{1991A&AS...91..199L}), and the variation of mesospheric sodium
column density is not negligible (e.g., \cite{Ge:1997xn}). These
factors substantially complicate any interpretation of the observed Na
lines.

\begin{figure*}[tbh]
 \begin{center} \FigureFile(160mm,220mm){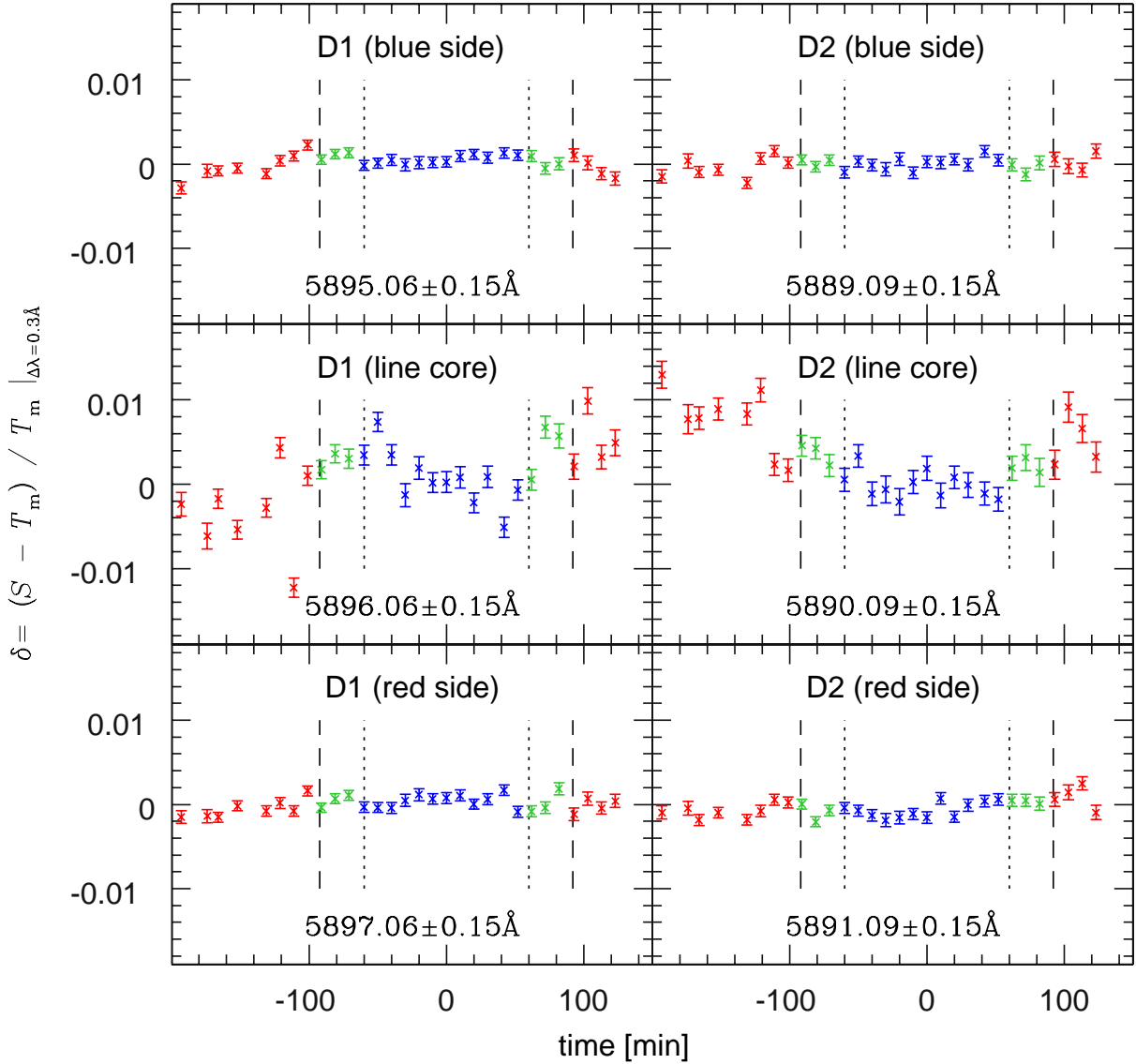} \caption{Difference 
  light curves around the sodium doublet lines for a
  smoothing width of $\Delta \lambda = 0.3$~\AA.  The error bars
  represent the Poisson noise $\sqrt{T_{\rm m}}$, and the horizontal
  axis indicates the time of each exposure with the origin at
  mid-transit. The time between the two dotted lines indicates the in-transit 
  phase, and those between the dashed line and dotted line show ingress
  (left side) and egress (right side) phases.
  The left side panels exhibit difference light curves
  for the Na D1 line at $5895.06 \pm 0.15$~\AA\ (top panel;
  continuum of blue side), $5896.06 \pm 0.15$~\AA\ (middle panel; line core),
  $5897.06 \pm 0.15$~\AA\ (bottom panel; continuum of red side). The right side
  panels are similar, but for the Na D2 line; $5889.09 \pm 0.15$~\AA\
  (top panel; continuum of blue side), $5890.09 \pm 0.15$~\AA\ (middle panel; 
  line core), $5891.09 \pm 0.15$~\AA\ (bottom panel; continuum of red side).}
  \label{Nadl015com}
  \end{center}
\end{figure*}
\begin{figure*}[ptbh]
 \begin{center}
  \FigureFile(160mm,95mm){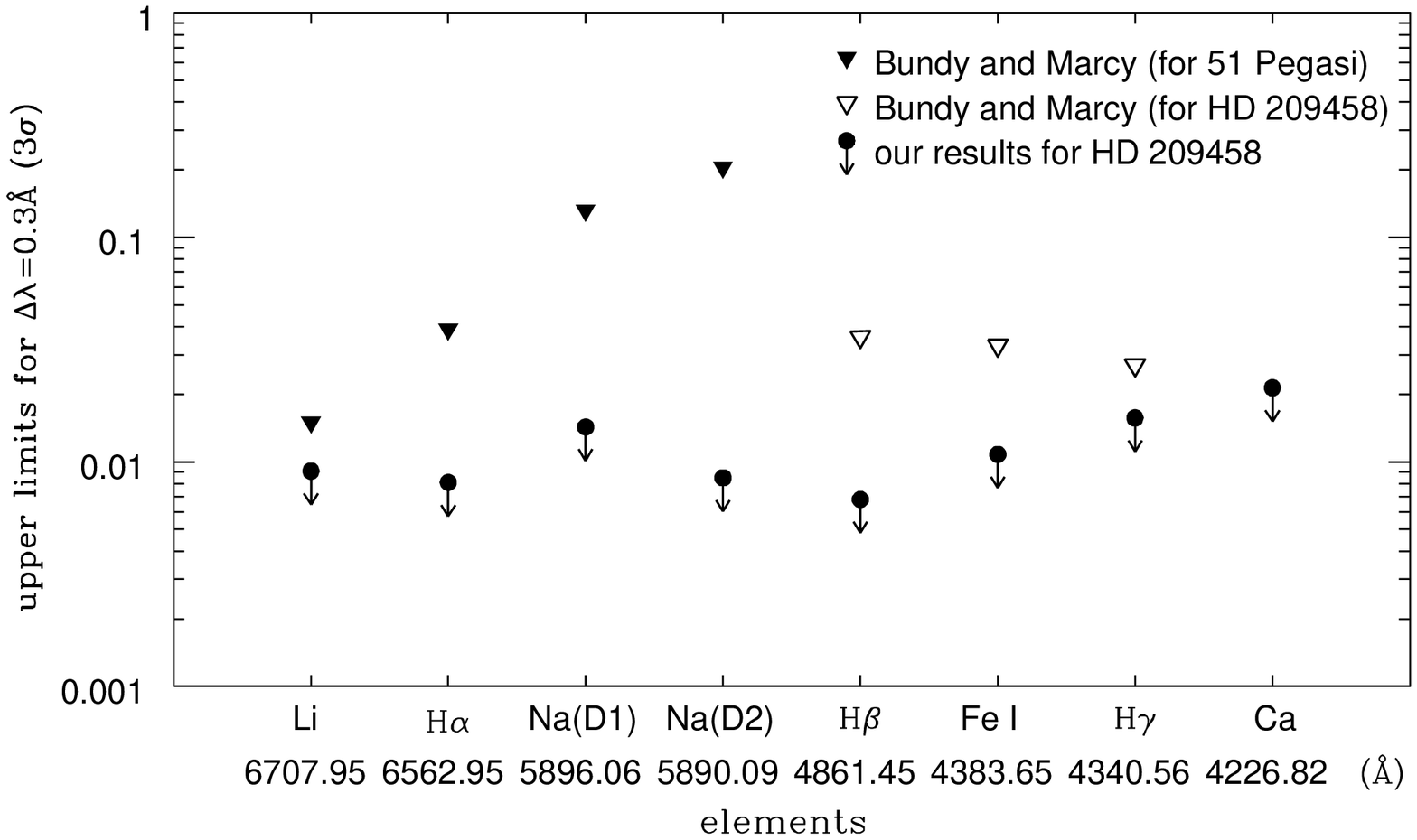} \caption{Comparison of the 
  Subaru HDS
upper limits for $\Delta \lambda = 0.3$~\AA\ with those of
\citet{2000PASP..112.1421B}. The vertical axis is fractional upper limits
on additional absorption during transit on a log scale.
The horizontal axis indicates atomic species and
corresponding line wavelength $\lambda_0$ in equation (1).
Note that \citet{2000PASP..112.1421B} did not report a Ca (4226.82~\AA)
result, and 
their upper limits on Li, H$\alpha$, and Na (D1, D2) are for 51 Pegasi.
\label{comparison}}
 \end{center}
%
 \begin{center}
  \FigureFile(160mm,160mm){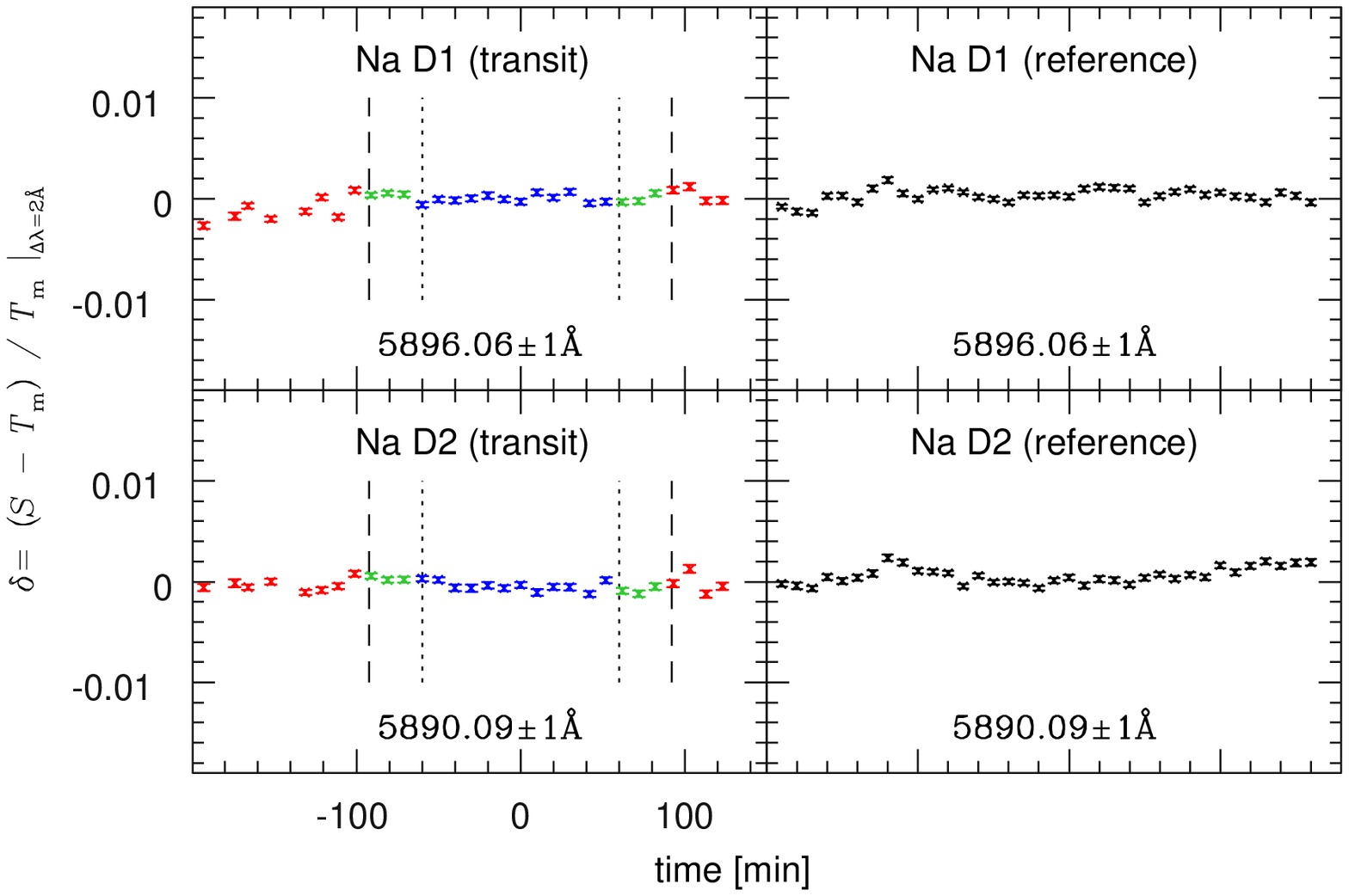} \caption{Difference light curves 
for $\Delta \lambda = 2$~\AA\ around the sodium doublet. The left panels
exhibit the light curves on the transiting night (the night of HST 2002 
October 24), and the right panels are those for the out-of-transit night
(October 26).
The vertical scale is matched to that of figure~\ref{Nadl015com}.
  \label{Nadl100}}
 \end{center}
\end{figure*}

There are time-dependent, coherent features with amplitudes that are
in excess of the Poisson noise level in the vicinity of the Na lines
(such features are also seen for other lines, but at a much reduced
level).  These anomalies strongly resemble those which would be
produced by a small, time-dependent wavelength shift. However,
measurements of the line center wavelengths (obtained by fitting a
Gaussian function to the line profile) of both stellar and telluric
features show them to be stable to within $\pm 0.01$~\AA\ (less than 1
pixel), and show no systematic drift with time. Thus, they do not
appear to be directly associated with errors or drift in our
wavelength calibration (determined from Thorium--Argon lamp spectra
obtained at the beginning and end of the HD~209458
exposures). Furthermore, these features are strongest {\it outside}
of the transit phase, especially at the beginning of the observation
when the target was near zenith, and thus when the instrument was rotating
most rapidly.
Hence, we believe the anomalies are unlikely to be due to real absorption
in the HD~209458 system.
Whatever their source, these features appear
to have almost exactly zero net photon count; they cancel out almost
completely when integrated over a smoothing width comparable to the
overall line width (e.g., 2~\AA). This suggests that they do not
compromise the achieved detection sensitivity for such a bandwidth.

\subsection{Comparison with Previous Results}

The same analysis was applied to the other selected lines.  For the
other lines, there was no significant interstellar absorption
contamination such as that observed for Na. In order to
compare our results with previous ones, we selected a bandwidth matched
to past studies.  In particular we followed the conventions of
\citet{2000PASP..112.1421B}, who performed transmission spectroscopy
for HD~209458 and 51 Pegasi in various atomic lines. They used $\Delta
\lambda \simeq 0.3$~\AA\ in order to emphasize the line cores.

The time variations of the fractional difference for $\Delta \lambda =
0.3$~\AA\ at the sodium doublet lines are plotted in
figure~\ref{Nadl015com}. It is evident that the difference light
curves for continuum regions are nearly consistent with the Poisson
noise of $\sim 0.07\%$, while those for line cores have a
significantly larger scatter than the expected Poisson noise of $\sim
0.15\%$. The results for line cores are $\delta_{\rm D1} = 0.12 \pm
0.48$\% and $\delta_{\rm D2} = - 0.70 \pm 0.28$\%, respectively.
Although $\delta_{\rm D2}$ indicates a 2.5$\sigma$ detection of
additional absorption, $\delta_{\rm D1}$ does not exhibit a
corresponding signal, and as noted at the end of the previous
subsection, the difference spectra have unexplained systematic
anomalies on a wavelength scale larger than the $0.3$~\AA\ smoothing
region. Thus, we do not regard this 2.5$\sigma$ signal as real
absorption, but rather as an indicator of systematic error. The
resulting upper limits are $3\sigma_{\rm D1} = 1.43$\% and
$3\sigma_{\rm D2} = 0.85$\%.

Figure~\ref{comparison} shows a comparison of our upper limits for HD~209458
with those of \citet{2000PASP..112.1421B} for HD~209458 (H$\beta$, H$\gamma$, 
Fe) and 51 Pegasi (Li, H$\alpha$, Na D1, and D2).
Note that although 51 Pegasi does not show transits, they consider
that the lack of transits does not rule out the possibility of excess
absorption features. 
The Subaru HDS upper bounds are the most stringent limits so far
that have been obtained with ground-based observations in the optical region.

\subsection{Results for Integration over the Full Line Width}

We repeated the same procedure described in the previous subsection
using a smoothing width $\Delta \lambda = 2$~\AA\ in order to cover the
typical width of the strong stellar absorption lines. In this case,
the contribution from the small systematic features discussed in
subsection 3.1 was nearly zero, because of the inversion symmetry of
those anomalies.  We display the difference light curves of the Na D
lines in figure~\ref{Nadl100}, as examples. The large scatter seen in
figure~\ref{Nadl015com} is almost entirely removed for this larger
smoothing width.  The results are $\delta_{\rm D1} = 0.06 \pm 0.09$\%
and $\delta_{\rm D1} = - 0.02 \pm 0.06$\%.  We also plot $\delta (t)$
for the out-of-transit data obtained on the night of HST 2002 October
26 for reference (right side).  These panels exhibit very similar
variations to those in the October 24 data, and thus support the
conclusion of no detected additional absorption during the transit
event.

Our fractional upper limits on additional absorption during transit for
all of the selected lines and both values of $\Delta \lambda$ lines are
given in table~\ref{summary}.  We did not set upper limits on Fe~{\sc
i} and Ca~{\sc i} in blue CCD, since the bands of $\Delta \lambda =
2$~\AA\ contain several irrelevant absorption lines.

\begin{figure}[htb]
 \begin{center}
  \FigureFile(72mm,72mm){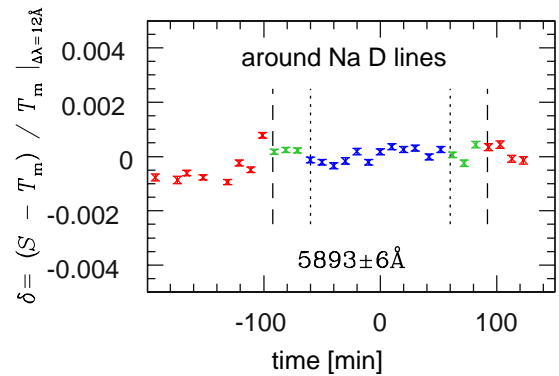} \caption{Difference light 
curve for $5893 \pm 6$~\AA, corresponding to the band in which
\citet{2002ApJ...568..377C} reported the detection of $0.02\%$ additional 
absorption.
\label{char}}
 \end{center}
\end{figure}

We also derived an upper limit on the Na D lines with $\Delta
\lambda = 12$~\AA, corresponding to the band in which
\citet{2002ApJ...568..377C} reported detection of additional absorption 
($5893 \pm 6$~\AA).  The resulting difference light curve is plotted in
figure~\ref{char}, and the result is $\delta_{\rm Na} = 0.03 \pm
0.04$\%.  Thus, we are not able to either confirm or contradict the
additional absorption of $0.02$\% reported by
\citet{2002ApJ...568..377C}, since it is beyond our sensitivity level.

\begin{table}[tbh]
\begin{center}
\caption{Summary of the 3$\sigma$ upper limits.}
\begin{tabular}{lccc}
\hline\hline
Element  & $\lambda_0$ & $\Delta \lambda = 0.3$~\AA & $\Delta \lambda = 2$~\AA\\
  & (~\AA) & (\%) & (\%)\\
\hline
\multicolumn{4}{c}{red CCD}\\
\hline
Li~{\sc i}& 6707.95 & 0.91 & 0.43\\
H$\alpha$& 6562.95 & 0.81 & 0.66\\
Fe~{\sc i}& 6024.20 & 0.84 & 0.18\\
Na~{\sc i}(D1)& 5896.06 & 1.43 & 0.28\\
Na~{\sc i}(D2)& 5890.09 & 0.85 & 0.18\\
\hline
\multicolumn{4}{c}{blue CCD}\\
\hline
H$\beta$& 4861.45 & 0.68 & 0.34\\
Fe~{\sc i}& 4383.65 & 1.08 & ---\\
H$\gamma$& 4340.56 & 1.57 & 0.88\\
Ca~{\sc i}& 4226.82 & 2.14 & ---\\
\hline
\end{tabular}
\label{summary}
\end{center}
\end{table}

\section{Evaluation of Systematic Errors}

\begin{figure*}[phtb]
 \begin{center}
  \FigureFile(160mm,120mm){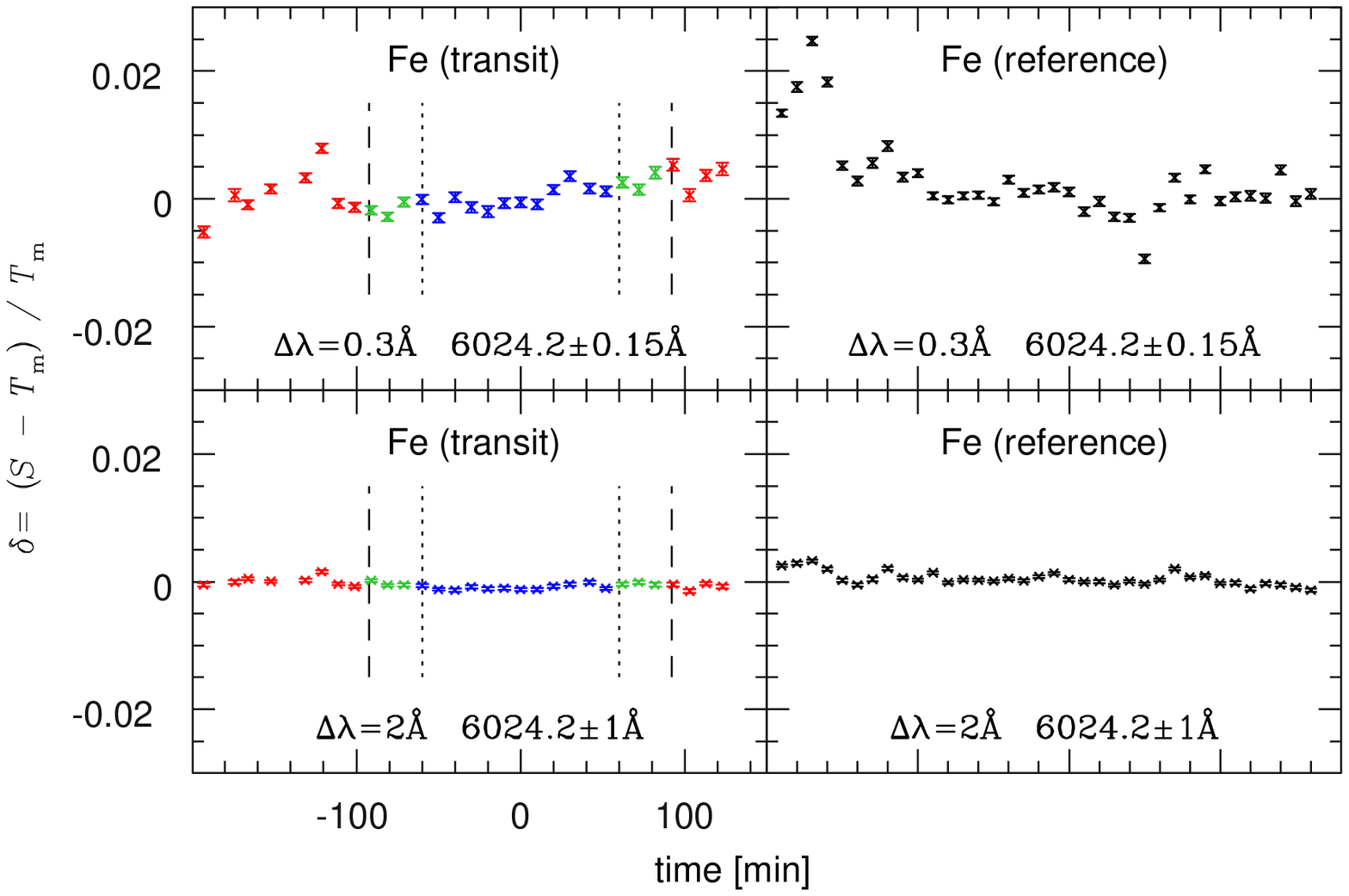} \caption{Difference light curves 
  for the reference Fe line at 6024~\AA. The upper panels show fractional differences of 
$\Delta \lambda = 0.3$~\AA, while the lower panels use $\Delta \lambda = 2$~\AA.
The standard deviations are 0.28\% (upper left), 0.65\% (upper right), 
0.06\% (lower left), and 0.10\% (lower right) respectively.}
  \label{fig5}
 \end{center}
%
 \begin{center}
  \FigureFile(160mm,105mm){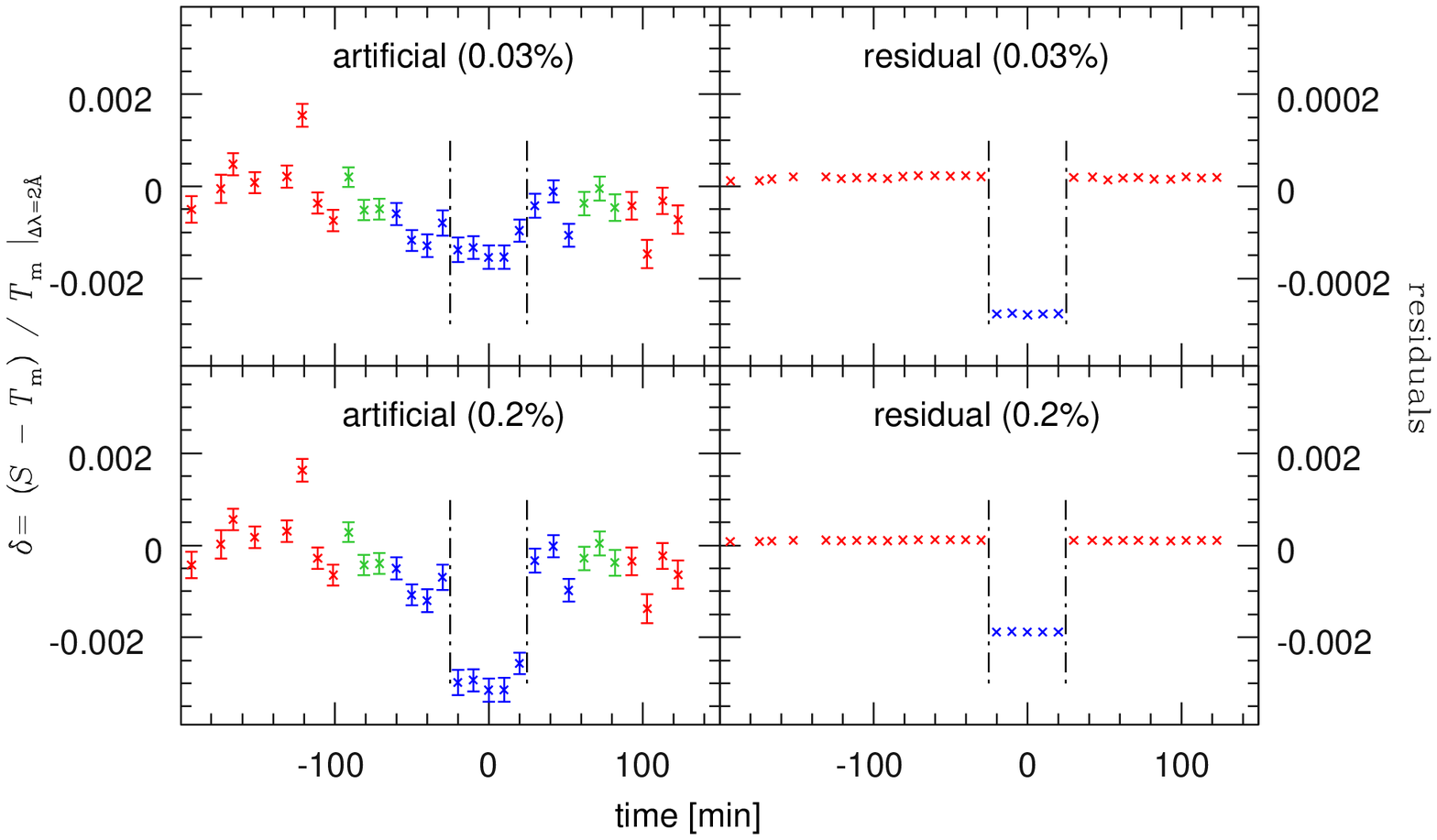} \caption{Difference light curves 
with artificial absorption (left panels).
The five data points between the dot-dashed lines indicate the results
with artificial signals.
The left panels show its recovery via our analysis procedure, and
the right panels show residuals that subtracted the original 
difference light curve (lower left panel of figure~\ref{fig5}.)
from the left panels.
   \label{art}}
 \end{center}
\end{figure*}

In order to evaluate the significance of high precision measurements,
such as those presented in this paper and to facilitate the planning
of similar, future observations, it is necessary to thoroughly
understand the sources and magnitudes of systematic errors.

We therefore carried out further analyses of the remaining systematic
effects, such as instrumental variations and telluric effects, which
may be serious obstacles to improving ground-based observations.
First of all, we checked the reliability of our correction method for
the instrumental variations and estimated the magnitude of any
remaining instrumental effects on the results.  Next, we confirmed
that the correction method does not dilute any real signal which is
above the Poisson noise limit.  Finally, we analyzed the correlation
between the spectra of HD~209458 and rapidly rotating star HD~42545 in
order to estimate the influence of terrestrial atmospheric effects.

\subsection{Estimation of Instrumental Effects}

Although our spectrum-matching procedure (Paper I gives its details)
is intended to correct for instrumental variations, the correction is
not perfect. To estimate the size of the remaining instrumental effects,
we selected an appropriate reference absorption line on the red CCD.
We adopted the Fe~{\sc i}] (intercombination) line at 6024.20~\AA~for the
following reasons: (i) there are no telluric lines within $\pm 2$~\AA~
\citep{1966sst..book.....M}, and (ii) the position of the line is near
the peak of the continuum, which minimizes the Poisson noise level.
The same procedure described in the previous section was applied with
$\Delta \lambda = 0.3$~\AA\ and $2$~\AA\ to the band around 6024.20~\AA.

The difference light curves are shown in figure~\ref{fig5}, and have
standard deviations of 0.28\% (upper left), 0.65\% (upper right),
0.06\% (lower left), and 0.10\% (lower right). As suggested by the
analysis presented in the previous sections, instrumental variations
appear to be removed approximately within the Poisson noise limit for
the larger value of $\Delta \lambda$ that integrates over the whole
line width. However, for the narrower band, which includes only the line
core, systematic effects of unknown origin increase the effective
noise to a level that is several times larger than the expected
Poisson noise.

\subsection{Recovery of Artificial Input Signals}

In order to verify that our reduction and analysis procedures do not
remove or dilute real signals, we injected an artificial signal of
0.03\% (corresponding to the Poisson noise) and 0.2\% into 5 out of 30
spectra in the October 24 data around $6024 \pm 1$~\AA, and analyzed
it via the same methods.  Figure~\ref{art} shows difference light
curves for $\Delta \lambda = 2$~\AA\ (left side) and residuals that
subtracted the original difference light curve from left panels (right
side).  We compared the mean and standard deviation of 25 realizations
without artificial injection ($\delta_{\rm no},\sigma_{\rm no}$), and
5 with artificial signals injected ($\delta_{\rm art}, \sigma_{\rm
art}$). We find that the residual $(\delta_{\rm no}-\delta_{\rm art}) \pm
(\sqrt{25\, \sigma_{\rm no}^2 /30 + 5\, \sigma_{\rm art}^2 /30})$ values are 
$0.0296 \pm 0.0003$\% (originally 0.03\%) and $0.1987 \pm 0.0009$\% 
(originally 0.2\%). We thus conclude that the procedures used in this paper 
can recover small input signals over the Poisson noise limits with
$\Delta \lambda = 2$~\AA\ quite satisfactorily and without significant
dilution.

\subsection{Telluric Effects}

Spectra obtained from the ground naturally contain many
telluric absorption features.  We obtained spectra of the rapidly
rotating star HD~42545 at the end of the night (airmass $\sim 1.9$)
and used this data to identify many such lines listed in
\citet{1966sst..book.....M}.  Many telluric water vapor and oxygen
absorption lines are visible in the spectra of the red CCD,
particularly.

\begin{figure}[htb]
 \begin{center}
  \FigureFile(80mm,130mm){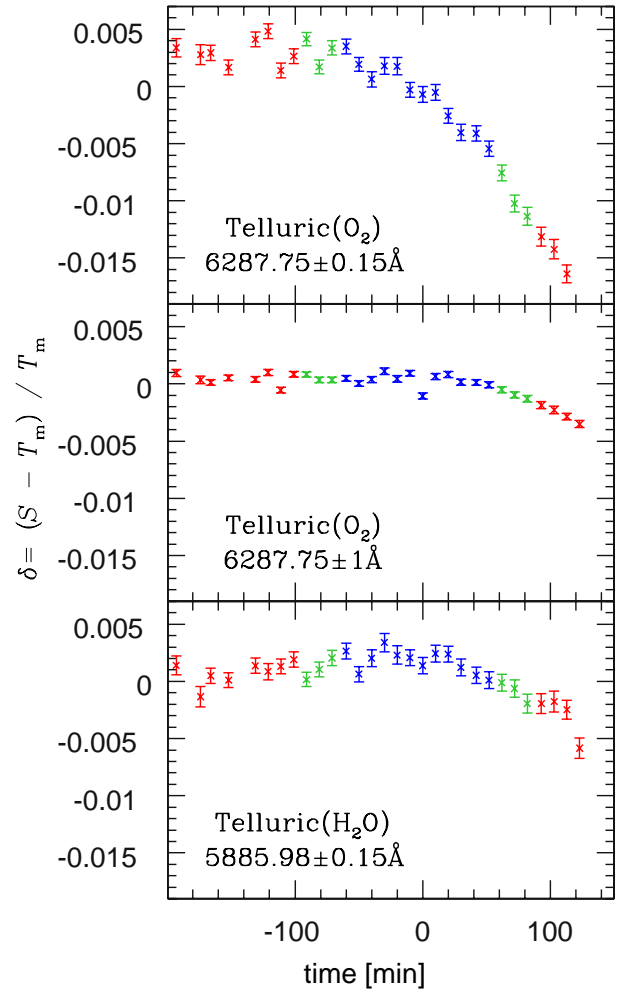} \caption{Difference light curves 
of a telluric oxygen line (upper and middle) and a water vapor 
feature (lower; near the Na D 
doublet). The strengths of both features increased near the end of the night
when the air mass became significantly larger with time.}
  \label{fig7}
 \end{center}
\end{figure}

The difference light curves for one strong telluric oxygen line at
$6287.75$~\AA~ with both $\Delta \lambda = 0.3$~\AA~(top panel) and
$\Delta \lambda = 2$~\AA~(middle panel) are shown in figure~\ref{fig7}.
The bottom panel shows the difference light curve
for $5885.98 \pm 0.15$~\AA, which contains one of the strongest
telluric water vapor lines near the Na D doublet, although it is 5 to
10 times weaker than the $6287.75$~\AA~oxygen line. These plots show
that the telluric features are measured to roughly within the Poisson
noise (as expected), and do not exhibit any anomalous behavior during the
transit phase, and vary most strongly at the end of the night when the
air mass became considerable ($\sim 2.0$).

Based on these results we estimate the telluric water vapor contamination
levels by comparison with the equivalent widths listed in
\citet{1966sst..book.....M}, and conclude that they were no more than 0.5\%
for $\Delta \lambda = 0.3$~\AA, and less than 0.1\% (3$\sigma$) for
$\Delta \lambda = 2$~\AA\ for all of the lines listed in
table~\ref{summary}.

\section{Summary and Conclusions}

An investigation of the optical spectra taken during a transit of
the HD~209458 system with Subaru HDS revealed no additional
absorption features associated with the transit.  The typical achieved
sensitivity level is $\sim 1$\% (3$\sigma$) for $\Delta \lambda =
0.3$~\AA\ (the line cores) in fractional difference spectra, and a few times
smaller for $\Delta \lambda = 2$~\AA (the full line width).
Although we cannot confirm the smaller level
of absorption in the Na D line that was observed from space
\citep{2002ApJ...568..377C}, our upper limits for the other optical 
transitions are consistent with, but more stringent than, those 
previously achieved from the ground.

Given the systematic effects that we encountered when working at very high
precision, one may wonder whether ground-based observations can ever be
competitive with space-based observations for this purpose. The answer is
unclear, but there are at least a few lines of future research that may
help to improve the prospects for ground-based observations. As we have
argued, the primary obstacle to higher-precision differential spectroscopy
seems to be small temporal variations in instrumental response; telluric
contamination is not negligible, but is removable to some extent.
Instruments that are purpose-built for increased stability (despite
telescope motion and instrument rotation) may be required. Alternatively,
it may be possible to use the multi-object capability of some
spectrographs to record simultaneous high-resolution, high-SNR spectra of
additional objects besides the exoplanetary system, and perform
differential spectrophotometry. Finally, an alternative method for probing
exoplanetary atmospheres was recently proposed by \citet{2004MNRAS.353L...1S}.
The idea is to measure the line-dependence of the Rossiter-McLaughlin effect
(e.g., \cite{1924ApJ....60...15R}; \cite{1924ApJ....60...22M};
\cite{2000A&A...359L..13Q}; \cite{2005ApJ...622.1118O}),
which is the spectral line distortion caused by the planet since it blocks 
a particular part of the rotation field of the face of the star.
This method involves a comparison of the shapes of different spectral 
lines, rather than their depths relative to the continuum.
Thus using this method in ground-based observations is quite reasonable,
because ground-based instruments would not exceed space-based ones in
photometric accuracy, but possibly could do so in spectral resolution.
We intend to use the existing Subaru/HDS data to investigate this
latter possibility.

\bigskip

We are very grateful to Kozo Sadakane for helpful advice concerning
telluric contamination. We also acknowledge close support of our
observations by Akito Tajitsu, who is a Support Astronomer of Subaru
HDS. This work is based on data from Subaru Telescope which is
operated by the National Astronomical Observatory of Japan, and
supported in part by a Grant-in-Aid for Scientific Research from the
Japan Society for Promotion of Science (No.14102004, 16340053), and by
NASA grant NAG5-13148. The visit of E.L.T at the University of Tokyo is 
supported by an invitation fellowship program for research in Japan from JSPS.
Work by J.N.W.is supported by NASA through
Hubble Fellowship grant HST-HF-01180.02-A, awarded by the Space
Telescope Science Institute, which is operated by the Association of
Universities for Research in Astronomy, Inc., for NASA, under contract
NAS~5-26555.
M.T. acknowledges support by Grant-in-Aids on
Priority Area No. 160777204 of the 
Ministry of Education, Culture, Sports, Science and Technology of Japan.
We wish to recognize and acknowledge the very
significant cultural role and reverence that the summit of Mauna Kea
has always had within the indigenous Hawaiian community.  We are most
fortunate to have had the opportunity to conduct observations from this
mountain.


\end{document}